# Genetic Algorithm in Audio Steganography

Manisha Rana[1], Rohit Tanwar[2]

[1]ECE Deptt, Manav Rachna International University
Faridabad, Haryana, India
[2]IT Deptt., Manav Rachna College of Engineering

*Abstract*--**With the advancement of communication technology, data is exchanged digitally over the network. At the other side the technology is also proven as a tool for unauthorized access to attackers. Thus the security of data to be transmitted digitally should get prime focus. Data hiding is the common approach to secure data. In steganography technique, the existence of data is concealed. GA is an emerging component of AI to provide suboptimal solutions. In this paper the use of GA in Steganography is explored to find future scope of research**.

*Keywords*-**HAS, SecretData , Genetic Algorithm, Mutation, Chromosome**

## I. INTRODUCTION

The word steganography comes from the Greek Steganos, which mean covered or secret and –graphy mean writing or drawing. Therefore, steganography means, literally, covered writing. Steganography is the art and science of hiding information such that its presence cannot be detected [7] and a communication is happening [8, 17]. A secret information is encoding in a manner such that the very existence of the information is concealed. Paired with existing communication methods, steganography can be used to carry out hidden exchanges. The main goal of steganography is to communicate securely in a completely undetectable manner [9] and to avoid drawing suspicion to the transmission of a hidden data [9]. It is not to keep others from knowing the hidden information, but it is to keep others from thinking that the information even exists. If a steganography method causes someone to suspect the carrier medium, then the method has failed.

There is a wide range of options that can act as cover file; Image, Audio, Text File, Video, etc. Audio file will be used as a cover media in the proposed work. The reason of focus towards audio is its "Masking Effect" and less work done in this area as compared to image steganography.

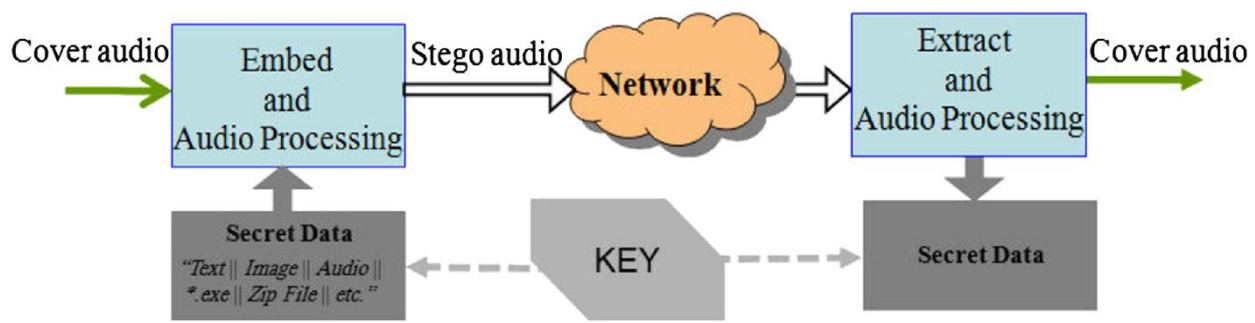

Fig.1. Audio Steganography

One common technique is based on manipulating the least-significant-bits (LSB) of every Audio frame by directly replacing the LSBs of the audio samples with the message bits. LSB methods typically achieve high capacity but unfortunately LSB insertion is vulnerable to attack as it is very easy to read the LSB's of every audio sample in sequence and to check whether it gives some meaningful message or not. Thus the security of the technique also lies in the sequence of accessing audio frames and hence target bits used for data hiding.

Genetic algorithm is a technique for optimization and search, which is based on the Darwinian principles of survival and reproduction (Goldberg 1989). The GA processes populations of chromosomes (individuals), which replace one population with another successively.

The chromosome in the GA is often held in binary encoding. Each chromosome represents a candidate solution in the searching space. The GA usually needs a fitness function to assign a score (fitness) to each chromosome in current population. The GA starts with initializing a population of individuals by guess. The individuals evolve through iterations, called generations. In each generation, each individual is evaluated against the fitness function. Genetic operators are used for individuals in the population to generate a next generation of individuals. The process is continued until some form of criterion is met (e.g., a given fitness is met) [3].

*STEP 1*: *Encrypt the secret message*

*STEP 2: Generate random population of size L (L=length of the Secret Message) with each individual member having n chromosomes (suitable solutions for the problem)*

*STEP 3: [Fitness] Evaluate the fitness f(x) of each chromosome individual in the population*

*STEP 4: [New population] Create a new population by repeating following steps until the newpopulation is complete*

   i. *[Selection] Select two parents from the population with the best fitness level (the better fitness, the bigger chance to be selected)*

   ii. *[Crossover] With a crossover probability, cross over the parents to form newoffspring (children). If no crossover was performed, offspring is an exact copy of parents.*

   iii. *[Mutation] With a mutation probability, mutate new offspring at each locus (position in chromosome).*

   iv. *[Accepting] Place new offspring in a new population*

*STEP 5: [Replace] Use new generated population for a further run of algorithm*

*STEP 6: [Test] If the end condition is satisfied, stop and return the best solution to current population*

*STEP 7: [Loop] Go to step 4 [5].*

A. Generating Population

Once the secret message is converted into its ASCII representative, the minimum and maximum values of the generated ASCII numbers are identified. The initial population will have a predetermined number of individuals which in this case is the total number of characters contained in the secret message. This population, which is a set of random numbers, is generated from the values that will fall between the identified minimum and maximum values. The individuals are grouped as a set of chromosomes containing genes. The two individuals with the highest fitness function will crossover to produce two offspring.

The two offspring will undergo mutation, then will be assigned a fitness value before re-introduction into the population. From the population the two least fit individuals will then be discarded, as the original population size needs to be maintained. This will continue until an optimal solution is obtained.

B. Fitness Function

To get the individuals that are most fit, set operators (i.e. Intersection A∩B) are used to compare the ASCII values (elements) that are in the secret message with those contained within the individuals. The more values (elements) of the secret message contained in an individual, the higher the fitness function.

C. Mutation

Mutation process is used to introduce scarce genes to the population. This is achieved by using the set operator (i.e. difference) A-B: elements in A that are not in B. B in this case is the union of all the values contained in all individuals in the population, and A is the values in the secret message. This is done to get the scarce genes, which will be introduced to the produced offspring. Randomly a gene in the produced offspring will be selected and substituted with the scarce gene/allele. Each offspring is mutated before introduction to the population. To ensure that the two least fit individuals are not discarded with genes that are needed for optimization, set difference operation is reapplied to get the scarce allele/gene, if any [5].

## II. MOTIVATION

Due to the advancement in IC Technology, most of the information is kept in electronic form. Consequently, the security of information has become a prime concern. Encrypting data has been the most popular approach to protect information but this protection can be broken with enough computational power. An alternate approach rather than encrypting data can be to hide it by making this information look like something else. In This way only friends would realize its true content. In particular, if the important data is hidden inside an image then everyone even your friends would view it as a picture. At the same time only your friends could retrieve the true information. This technique is often called data hiding or steganography. Besides cryptography, steganography can also be used to secure information.

Steganography is a technique of hiding information (in digital media these days). In contrast to cryptography, the message or encrypted message is embedded in a digital host before passing it through the network, thus the existence of the message is unknown. Besides hiding data for confidentiality, this approach of information hiding can be extended to copyright protection for digital media: audio, video, and images. The growing possibilities of modern communications need the special means of security especially on computer network. The network security is becoming more important as the number of data being exchanged on the Internet is increasing. Therefore, the confidentiality and data integrity are required to be protected against unauthorized access and use. This has resulted in an explosive growth in the field of information hiding[1].

Until recently, information hiding techniques received very less attention from the research community and from industry than cryptography. This situation is, however, changing rapidly and the first academic conference on this topic was organized in 1996. There has been a rapid growth of interest in steganography for two main reasons [6].

(i) The publishing and broadcasting industries have become interested in techniques for hiding encrypted copyright marks and serial numbers in digital films, audio recordings, books and multimedia products.

(ii) Moves by various governments to restrict the availability of encryption services have motivated people to study methods by which private messages can be embedded in seemingly innocuous cover messages.

In recent years, many successful steganography methods have been proposed. Among all the methods, LSB (Least Significant Bit) replacement method is widely used due to its simplicity and large capacity. The majority of LSB steganography algorithms embed messages in Image in spatial domain, such as BPCS, PVD. Some others, such as Jsteg, F5, Outguess, embed messages in DCT frequency domain (i.e. JPEG images). In the LSB methods for audio steganography, secret message is converted into binary string. Then the least significant bit-plane of audio samples is replaced by the binary string. The LSB embedding achieves good balance between the payload capacity and audio quality. The main advantage of this method is the Masking Effect of HAS(low strength voices are suppressed by those with comparative higher strength); however unfortunately, it is extremely vulnerable to attacks, such as audio compression and format conversion [2].

The LSB method has low robustness against attacks that try to reveal the hidden message. Since this technique usually modify the bits of lowest layer in the samples -LSBs, it is easy to reveal the hidden message if the low transparency causes suspicious.

To overcome this problem, there came the concept of Randomized LSB; where the next bit to be used for hiding message bits is selected based on some pseudo random technique. Since the ordering in which the target bit for substitution is selected is not obvious or known earlier, it is difficult for the attacker to reveal the hidden message until the pseudo random code is known.

The LSB method has low robustness against distortions with high average power. Unintentional attacks like transition distortions could destroy the hidden message if it is embedded in the bits of lower layers in the samples -LSBs.

To overcome this problem, the message can be hidden in other layers of bits than the LSB. However the bits in upper layers contribute a significant amount of audio relevant details and the bits in lower layers contribute more fine details. Thus the selection of bits to be used for replacing message bits is to be done keeping these points in mind.

The genetic algorithm (GA) has been used plentifully in information hiding discipline these years (Chu et al. 2008; Huang et al. 2007; Pan et al. 2004) and has been shown to
be an effective technique for improving the performance of information hiding systems[3]. Genetic algorithms are used as an aiding tool for generating and optimizing security protocol (Zarza *et al*., 2007). Using Genetic Algorithms that are based on the mechanism of natural genetics and the theory of evolution, we can design a general method to guide the steganography process to the best position for data hiding.

III. LITERATURE REVIEW

K.P.Adhiya and Swati A. Patil propose a steganographic method for embedding textual information in audio. In this method each audio sample is converted into bits and then the textual information is embedded in it. The last 4 bits of this binary is taken into consideration and applying redundancy of the binary code the prefix either 0 or 1 is used.
*Advantage*: 16bitWAV and 8bitWAVaudio file are supported and the secret message can be hidden in the audio file with less storage capacity.
*Disadvantage*: The proposed algorithm gives better result for 16 bit wav audio as compared to 8 bit .

Prof. Samir Kumar Bandyopadhyay and Barnali Gupta Banik gives an overview of two primitive techniques to get an idea of how steganography in audio file works. LSB modification and phase encoding technique are very primitive in steganography. An effective audio steganographic scheme should possess the following three characteristics: Inaudibility of distortion, Data Rate and Robustness. These characteristics are called the magic triangle for data hiding.
*Advantage*: This method is easy to implement but is very susceptible to data
*Disadvantage*: This method can be used when only a small amount of data needs to be concealed.

Jayaram P, Ranganatha H R, Anupama H S discuss different types of audio steganographic methods, advantages and disadvantages.
*Advantage*: This paper concludes that audio data hiding techniques can be used for a number of purposes other than covert communication or deniable data storage, information tracing and finger printing, tamper detection.

R SRIDEVI, DR. A DAMODARAM and DR. SVL.NARASIMHAM gives basic idea behind to provide a good, efficient method for hiding the data from hackers and sent to the destination in a safer manner.
*Advantage*: This proposed system is to provide an efficient method for hiding the data from hackers and sent to the destination in a safe manner and this system will not change the size of the file even after encoding and also suitable for any type of audio file format.
*Disadvantage*: The quality of sound depends on the size of the audio which the user selects and length of the message.

Samir Kumar Bandyopadhyay et. Al 2010] detects certain flaws in mostly substitution techniques of

steganography:

1) *Having low robustness against attacks which try to reveal the hidden message. Since substitution techniques usually modify the bits of lower layers in the samples -LSBs, it is easy to reveal the hidden message if the low transparency causes suspicious*

2) *Having low robustness against distortions withhigh average power. Unintentional attacks like transition distortions could destroy the hidden message if is embedded in the bits of lower layers in the samples –LSBs[4].*

Christine K. Mulunda,Peter W. Wagacha, Alfayo O. Adede worked with text as the cover medium with the aim of increasing robustness and capacity of hidden data. Elitism was used for the fitness function.

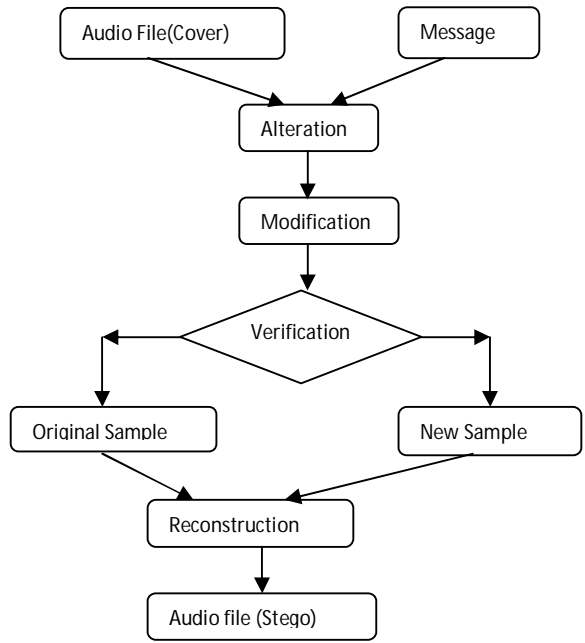

Fig. 2 : Stenography

*Advantages:* this approach works, achieving effective optimization, security, and robustness.

*Disadvantage:* Applicable for text files only.

IV. GENETIC ALGORITHM IN STEGANOGRAPHY

As Fig. 1 shows, there are four main steps in this algorithm that are explained below.

*A. Alteration*

At the first step, message bits substitute with the target bits of samples. Target bits are those bits which place at the layer that we want to alter. This is done by a simple substitution that does not need adjustability of result be measured.

*B. Modification*

In fact this step is the most important and essential part of algorithm. All results and achievements that we expect are depending on this step. Efficient and intelligent algorithms are useful here. In this stage algorithm tries to decrease the amount of error and improve the transparency. For doing this stage, two different algorithms will be used.

One of them that is more simple likes to ordinary techniques, but in aspect of perspicacity will be more efficient to modify the bits of samples better. Since transparency is simply the difference between original sample and modified sample, with a more intelligent algorithm, I will try to modify and adjust more bits and samples than some previous algorithms. If we can decrease the difference of them, transparency will be improved. There are two example of adjusting for expected intelligent algorithm below.

Sample bits are: 00101111 = 47

Target layer is 5, and message bit is 1

Without adjusting: 00111111 = 63 (difference is 16)

After adjusting: 00110000 = 48 (difference will be 1 for 1 bit embedding)

Sample bits are: 00100111 = 39

Target layers are 4&5, and message bits are 11

Without adjusting: 00111111 = 63 (difference is 24)

After adjusting: 00011111 = 31 (difference will be 8 for 2 bits embedding)

Another one is a Genetic Algorithm which the sample is like a chromosome and each bit of sample is like a gene. First generation or first parents consist of original sample and altered sampled. Fitness may be determined by a function which calculates the error. It is clear, the most transparent sample pattern should be measured fittest. It must be considered that in crossover and mutation the place of target bit should not be changed.

*C. Verification*

In fact this stage is quality controller. What the algorithm could do has been done, and now the outcome must be verified. If the difference between original sample and new sample is acceptable and reasonable, the new sample will be accepted; otherwise it will be rejected and original sample will be used in reconstructing the new audio file instead of that

*D. Reconstruction*

The last step is new audio file (stego file) creation. This is done sample by sample. There are two states at the input of this step. Either modified sample is input or the original sample that is the same with host audio file. It is why we can claim the algorithm does not alter all samples or predictable samples. That means whether which sample will be used and modified is depending on the status of samples (Environment) and the decision of intelligent algorithm.

## V. CONCLUSION

Unlike cryptography, Steganography conceals the existence of the secret message in cover audio file. Secrecy of the data hiding using steganography lies in the process used for embedding message in audio file. Substitution is generally used technique in implementing steganography that provides high data hiding capacity but its limitation lies in its simplicity of embedding process making it less robust. GA can be used to increase the robustness of substitution techniques while maintaining the significant data hiding capacity. This can be achieved by exploring the main beauty of GA,"Survival of the fittest".

## VI. REFRENCES